\hsize 12.5truecm 
\vsize 19truecm 
\hfuzz10000pt

\noindent
{SCHR\"{O}DINGER'S  INTERPOLATION PROBLEM AND ITS PROBABILISTIC\par
SOLUTIONS}\hfill \break
\medskip
\noindent
{Piotr Garbaczewski}\hfill \break
\noindent
{Institute of Theoretical Physics, University of Wroc{\l}aw,}
\hfill \break
\noindent
{PL-50 204 Wroc{\l}aw, Poland}
\hfill \break
\noindent
{e-mail address: pgar@ift.uni.wroc.pl}
\hfill \break
\bigskip

{\bf Abstract:}
Probablistic solutions of the so called Schr\"{o}dinger boundary data
problem provide for a unique Markovian interpolation between any two
strictly positive probability densities designed to form the
input-output statistics data for a certain dynamical process
taking place in a
finite-time interval. The key problem is to select the jointly
continuous in all variables positive semigroup  kernel, appropriate
for the phenomenological (physical) situation.
\medskip

The issue of {\it deriving } a microscopic dynamics  
from the  input-output statistics
data (analyzed in terms of densities) was addressed,
as the Schr\"{o}dinger problem of a probabilistic interpolation,
in a number of publications
[1]-- [6].
We shall consider Markovian propagation scenarios so remaining within
the well established framework, where for any two Borel sets
$A,B\subset R$ on which
the strictly positive boundary  densities $\rho (x,0)$
and $\rho (x,T)$ are defined, the transition probability
$m(A,B)$ from the set $A$ to the set $B$ in
the time interval $T>0$ has a bi-variate density given in
a specific factorized
form:  $m(x,y)=f(x)k(x,0,y,T)g(y)$, with marginals:
$${\int  m(x,y) dy =\rho (x,0)\, ,\, \int m(x,y)dx =\rho (y,T)}
\eqno (1)$$
Here, $f(x), g(y)$ are the a priori unknown functions, to come out as
solutions of the integral  system of equations (1),
provided that in addition to the density boundary data we have in
hands  {\it any}  strictly positive, continuous  in space
variables {\it function }  $k(x,0,y,T)$.
Additionally, we impose a restriction that $k(x,0,y,T)$
 represents  a  certain strongly
continuous dynamical semigroup  kernel, while given at the time
interval borders:  it  secures  the
Markov property of the sought for stochastic process.

It is the major mathematical discovery 
[2] that,
without the semigroup 
assumption  {\it but}  with the prescribed, nonzero boundary data 
$\rho (x,0),\rho (y,T)$ \it and \rm with the  strictly positive
continuous function
$k(y,0,x,T)$, the  system (1) of integral equations
admits a unique solution in terms of two nonzero, locally integrable 
 functions $f(x), g(y)$ of the same
sign (positive, everything is up to a multiplicative constant).

If $k(y,0,x,T)$ is a particular, confined to the time interval
endpoints,  form of a concrete semigroup kernel
$k(y,s,x,t), 0\leq s\leq t<T$,
then there exists  a transition density:
$${p(y,s,x,t)=k(y,s,x,t){{\theta (x,t)}\over {\theta (y,s)}}}
\eqno (2)$$
defined in terms of functions
$${\theta (x,t)=\int dy k(x,t,y,T)g(y)}, 
\qquad\theta _*(y,s)=\int dx k(x,0,y,s)f(x) \eqno (3)$$
which implements a consistent Markovian propagation of the
probability density
$\rho (x,t)= \allowbreak 
\theta (x,t)\allowbreak \theta _*(x,t)$ between its boundary versions,  
according to the standard transport recipe:
${\rho (x,t) = \int p(y,s,x,t)\rho (y,s)dy}$.
For a given semigroup which is characterized by its generator
 (Hamiltonian), the kernel $k(y,s,x,t)$ and the emerging transition
 probability density
$p(y,s,x,t)$  are unique in view of the uniqueness of solutions
$f(x),g(y)$  of (1). For Markov processes, the knowledge of the
transition probability density $p(y,s,x,t)$ for all intermediate
times $0\leq s< t\leq T$  suffices
for the derivation of all other relevant characteristics. 

In the framework of the Schr\"{o}dinger problem the choice of the
integral  kernel $k(y,0,x,T)$ is arbitrary,
except for the strict positivity and
continuity demand.
It is thus rather natural to ask for the most general
stochastic interpolation, that is admitted under the above premises.

Clearly, the familiar  strictly positive (Feynman-Kac) semigroup
kernels generated by Laplacians plus suitable potentials are very special 
examples in 
a surprisingly rich encompassing family. Indeed, the concept of the
"free noise", normally characterized by a Gaussian probability 
distribution 
appropriate to a Wiener process, can be extended to all infinitely 
divisible 
probability laws via the L\'{e}vy-Khintchine formula
[7].
It expands our framework from continuous diffusion processes to  jump 
or combined diffusion--jump propagation scenarios. All such (L\'{e}vy) 
processes 
are associated with strictly positive dynamical semigroup kernels, and 
all of them give rise to Markov solutions of the Schr\"{o}dinger 
stochastic interpolation problem (1)-(3).

At this point, let us remark that apart from the wealth of physical
phenomena described in terms
of Gaussian stochastic processes, there is a number of physical 
problems where 
the Gaussian tool-box proves to be insufficient to provide satisfactory 
probabilistic explanations. Non--Gaussian L\'{e}vy processes 
naturally 
appear in the study of transient random walks when long-tailed 
distributions 
arise 
[8,9].
They are also found necessary to analyze fractal
random walks 
[10], intermittency phenomena, anomalous diffusions, and 
turbulence at high Reynolds numbers 
[8,11].
On the other hand, our formulation  of the Schr\"{o}dinger interpolating
dynamics can be regarded as a straightforward inversion of the well
developed programme of studying dynamical systems (chaotic included) in
terms of densities, 
[12,13].

Let us consider  Hamiltonians (semigroup generators)
of the form $H=F(\hat{p})$, where
$\hat{p}=-i
\nabla $ stands for the momentum operator and  for 
$-\infty <k<+\infty $, $F=F(k)$ is a real valued, 
bounded from below, locally integrable function. Then, 
$exp(-tH)=\int_{-\infty }^ {+\infty } exp[-tF(k)] dE(k) $, $t\geq 0$, 
where $dE(k)$ is the spectral measure of $\hat{p}$.  
We simplify further  discussion  by considering  processes in one
spatial dimension. Because  $(E(k)f)(x)=
{1\over {\sqrt {2\pi }}}\int_{-\infty }^{k} exp(ipx) \hat{f}(p) dp $, 
where  $\hat{f}$
is the Fourier transform of $f$, we learn that 
$$[exp(-tH)]f(x) = [\int_{-\infty }^{+\infty } exp(-tF(k)) dE(k)f](x) 
=  $$
$${{1\over {\sqrt {2\pi }}}\int_{-\infty }^{+\infty } exp(-tF(k)) exp(ikx)
\hat{f}(k) dk = [exp(-tF(p)) \hat{f}(p)]^{\vee }(x)}\eqno (4)$$
where the superscript $\vee $ denotes the inverse Fourier transform.
 
Let us set $k_t={1\over {\sqrt {2\pi }}}[exp(-tF(p)]^{\vee }$, then the 
action of $exp(-tH)$ can be given in terms of a convolution: 
$exp(-tH)f = f*k_t$, where $(f*g)(x): =\int_R g(x-z)f(z)dz $. 

 We are interested in those $F(p)$ which give rise to
 positivity preserving semigroups:  if $F(p)$ satisfies the celebrated 
 L\'{e}vy-Khintchine formula, then $k_t$ is a positive measure for all 
 $t\geq 0$.
The most general case refers to a contribution from three types of 
processes:  deterministic, Gaussian, and an exclusively jump process.
Let us concentrate on the integral part of the L\'{e}vy-Khintchine formula,
which is responsible for arbitrary stochastic jump features:
$${F(p) = -  \int_{-\infty }^{+\infty } [exp(ipy) - 1 - 
{ipy\over {1+y^2}}] 
\nu (dy)}\eqno (5)$$
where $\nu (dy)$ stands for the so-called L\'{e}vy measure.
The disregarded Gaussian contribution would read $F(p)=p^2/2$; cf. Refs.
[1-6]
 for an exhaustive discussion of  related topics based on the concept
of the conditional Wiener measure.

There are not many explicit examples (analytic formulas for probability
densities involved) for processes governed  by (5), except possibly for
the so called stable probability laws. The best known example is the
classic Cauchy density.
Let us focus our  attention on two selected choices  for the
characteristic exponent $F(p)$, namely: 
$F_0(p)=|p|$ which is the Cauchy process generator, and
$F_m(p) =\sqrt {p^2 + m^2} - m, m>0$. Here, we have
chosen suitable units so as to eliminate inessential parameters. 
The latter exponent is another form of the familiar  classical
relativistic Hamiltonian, better known as
$\sqrt {m^2c^4+c^2p^2 }-mc^2$ where $c$ is the velocity of light.
The respective semigroup generators  $H_0, \, H_m$ are
pseudodifferential operators. The associated kernels $k^0_t,\, k^m_t$
in view 
of the "free noise"restriction (no potentials at the moment)
 are transition densities of
the jump (L\'{e}vy) processes with intensities  regulated by
the  corresponding  L\'{e}vy
measures $\nu _0(dy),\, \nu _m(dy)$.   
The affiliated  Markov processes solving the Schr\"{o}dinger problem 
(1)-(3) immediately follow.
It is instructive to notice that like in   case of more traditional
Gaussian derivations
[4], the identities  $\theta (x,t)\equiv  1,\,
\theta _*(x,t):= \overline {\rho }(x,t)$ 
imply the pseudodifferential analog of
the Fokker-Planck equation. It is a consequence of
$[exp(-tH)\overline {\rho }](x)=\overline {\rho }(x,t)$ and  of
the identification $F(p\rightarrow -i\nabla ):=H$. For
example there holds:
${F_0(p)\Longrightarrow \partial _t{\overline {\rho }}(x,t)=
 - |\nabla|{\overline {\rho }}(x,t)}$. This evolution rule gives rise to
 the Cauchy process with its  long-tailed probability density  $rho
 (x,t)={1\over \pi }{t\over {t^2+x^2}}$ and the transition
 probability density (e.g.
 the semigroup kernel in this free propagation case)  of the same
 functional form with $x\rightarrow x-y$ and $t\rightarrow t-s$.
 Let us emphasize that the existence and uniqueness of
 solutions proof for the Schr\"{o}dinger problem  extends to all cases
 governed by the infinitely divisible probability laws, and can be
 generalized to encompass the additive perturbations by  physical
 potentials (in analogy with the familiar Feynman-Kac formula).

Our     semigroups are holomorphic, hence we can replace
the time parameter 
$t$ by a complex one $\sigma =t+is, \,t>0$  so that $exp(-\sigma H)= 
\int_R
exp(-\sigma F(k))\, dE(k)$.  Its  action is defined 
by
 ${[exp(-\sigma H)]f = [(\hat{f}exp(-\sigma F)]^{\vee } = f*k_{\sigma }}
$. Here, the kernel reads $k_{\sigma }={1
\over {\sqrt {2\pi }}}[exp(-\sigma F)]^{\vee }$. Since $H$ is 
selfadjoint, the 
limit $t\downarrow 0$ leaves us with the unitary group $exp(-isH)$, 
acting in 
the same way: $[exp(-isH)]f = [\hat{f} exp(-isF)]^{\vee }$, except 
that now 
$k_{is}: = {1\over {\sqrt {2\pi}}}[exp(-isF)]^{\vee }$ in general is 
\it not  
\rm a measure. In view of  unitarity, the unit ball in $L^2$ is an 
invariant of the dynamics. Hence density measures can be associated with 
solutions 
of the Schr\"{o}dinger pseudodiferential equations:
${F_0(p)\Longrightarrow i\partial _t \psi (x,t) =  |\nabla | \psi (x,t)}
$ or
${F_m(p)\Longrightarrow i\partial _t\psi (x,t) =
[\sqrt {-\triangle + m^2} 
- m ]\psi (x,t)}$, if
provided with the appropriate initial data functions $\psi (x,0)$.
Let us point out that we know in
detail how the analytic continuation in time of the Laplacian generated 
holomorpic semigroup induces a mapping to diffusion processes
of a quantum mechanical proveninence
(since the standard Schr\"{o}dinger equation  $i\partial _t\psi (x,t)=
-\triangle \psi (x,t)$ is involved).

All that ultimately submits the unitary (quantum) Schr\"{o}dinger
picture dynamics, with quite a variety of admissible semigroup generators
to be used instead of  the traditional Laplacian, to the stochastic
 analysis in the framework of the
Schr\"{o}dinger (again) boundary data problem. The natural
question to be answered is:
 what are the stochastic processes consistent with the probability
measure 
dynamics $\rho (x,t)=|\psi (x,t)|^2$  determined by  pseudodifferential 
equations, eventually in the presence of external force
fields ?
The answer to this and related questions of the more pedestrian,
nonequlibrium statistical physics  origin can be found elswhere,
[6,13,14].

\vskip 14pt

{\bf Acknowledgement}: The author receives support from
the  KBN research grant No 2 P302 057 07.

\vfill
\eject
\centerline{REFERENCES}

\item{[1]} E. Schr\"{o}dinger, 
{\it Ann. Inst. Henri Poincar\'{e}}, {\bf  2}, 269
(1932).

\item{[2]} B. Jamison, 
{\it Z. Wahrsch.  verw. Geb.} {\bf 30}, 65 (1974).

\item{[3]} J.C. Zambrini, 
{\it J. Math. Phys.}  {\bf 27}, 2307 (1986).

\item{[4]} Ph. Blanchard, P. Garbaczewski, 
{\it Phys. Rev.}  {\bf E49}, 3815
(1994).

\item{[5]} P. Garbaczewski, R. Olkiewicz, 
{\it Phys. Rev.}  {\bf A51}, 3445 (1995).

\item{[6]} P.Garbaczewski, J. R. Klauder, R. Olkiewicz, 
{\it Phys. Rev.} {\bf E51}, 4114 (1995).

\item{[7]} M. Reed, B.Simon, 
{\it Methods of Modern Mathematical Physics}, 
Academic, New York 1978, vol. IV.

\item{[8]} E.W. Montroll, B.J. West., in: 
{\it Fluctuation Phenomena}, 
ed. by E. W. Montroll and J. L. Lebowitz, North-Holland, Amsterdam, 1987.

\item{[9]} H.C. Fogedby, 
{\it Phys. Rev. Lett.}, {\bf 73}, 2517 (1994).

\item{[10]} B.B. Mandelbrot, 
{\it The Fractal Geometry of Nature}, 
W. H. Freeman, New York 1982.

\item{[11]} J. Klafter, A. Blumen, M. F. Shlesinger, 
{\it Phys. Rev.} {\bf A35}, 3081 (1987).

\item{[12]} A. Lasota, M. C. Mackey, 
{\it Chaos, Fractals, and Noise},
Springer-Verlag, Berlin 1994.

\item{[13]} P. Garbaczewski, M. Wolf, A. Weron (eds.), 
{\it Chaos-The Interplay Between Stochastic and Deterministic Behaviour}, 
Karpacz'95 Proc., Springer-Verlag, Berlin 1995.

\item{[14]} P. Garbaczewski, R. Olkiewicz, 
{\it Feynman-Kac Kernels in
Markovian Representations of the Schr\"{o}dinger 
Interpolating Dynamics}, subm. forpubl.

\bye

\end{document}